\documentclass{vgtc}                          

\ifpdf
  \pdfoutput=1\relax                   
  \usepackage{graphicx}                
  \DeclareGraphicsExtensions{.pdf,.png,.jpg,.jpeg} 
\else
  \usepackage{graphicx}                
  \DeclareGraphicsExtensions{.eps}     
\fi%

\graphicspath{{figures/}{pictures/}{images/}{./}} 

\usepackage{microtype}                 
\PassOptionsToPackage{warn}{textcomp}  
\usepackage{textcomp}                  
\usepackage{mathptmx}                  
\usepackage{times}                     
\usepackage{cite}                      
\usepackage{tabu}                      
\usepackage{booktabs}                  

\onlineid{1109}

\vgtccategory{Research}

\vgtcinsertpkg

\title{A User-centered Design Study\\ in Scientific Visualization Targeting Domain Experts}

\author{Yucong (Chris) Ye\thanks{e-mail: chrisyeshi@gmail.com}\\ %
        \scriptsize UC Davis %
\and  Franz Sauer\thanks{e-mail: fasauer@ucdavis.edu}\\ %
     \scriptsize UC Davis %
\and Kwan-Liu Ma\thanks{e-mail: ma@cs.ucdavis.edu}\\ %
     \scriptsize UC Davis %
\and Konduri Aditya\thanks{e-mail: akondur@sandia.gov}\\ %
     \scriptsize Sandia National Laboratories %
\and Jacqueline Chen\thanks{e-mail:  jhchen@sandia.gov}\\ %
     \scriptsize Sandia National Laboratories}

\abstract{The development and design of visualization solutions that are truly usable is essential for ensuring both their adoption and effectiveness. User-centered design principles, which focus on involving users throughout the entire development process, are well suited for visualization and have been shown to be effective in numerous information visualization endeavors.
In this paper, we report a two year long  collaboration with combustion scientists that, by applying these design principles,  generated multiple results including an in situ visualization technique and a post hoc probability distribution function (PDF) exploration tool. Furthermore, we examine the importance of user-centered design principles and describe lessons learned over the design process in an effort to aid others who also seek to work with scientists for developing effective and usable scientific visualization solutions.} 


\usepackage{subcaption}

\begin{document}


\firstsection{Introduction}

\maketitle



Developing a truly usable software tool requires a great deal of effort to formulate the design based on users' needs. Without sufficient communication, developers run the risk of misinterpreting the needs and goals of their end users, thus resulting in a less effective tool. To avoid such an outcome, developers should adopt a user-centered design process, 
which focuses on involving the end users throughout the entire process as much as possible in order to effectively meet their needs. User-centered design may be guided by several key principles~\cite{noauthor_2017-cx}: an explicit understanding of users and context, user-centered evaluation, addressing the whole user experience, involving users throughout iterative design and development, and a team that includes multidisciplinary skills and perspectives.

In the field of visualization, user-centered design has been extensively discussed 
and employed~\cite{Koh2011-vr,Slocum2003-fa,Lloyd2011-yz}, but it is more often done 
for the development of visualization applications serving a large user group 
such as the public. In this case, usability studies are generally easier to arrange 
because of having wider choices of human subjects or 
the option to use Mechanical Turk~\cite{MTurk}. Considering the development of 
a visualization solution for a small group of domain experts such as 
scientists, surprisingly, we found very little has been done to involve 
the users throughout the full design and evaluation process, i.e., 
the practice of user-centered design. Based on our own experience in working with
scientists, we believe one reason has to do with the level of effort required 
to communicate and understand the sophisticated mathematical and physical 
underpinning of what the scientists want to visualize. 
It's not always possible to obtain scientists' commitment to participate
in a long, iterative design process, and most academic visualization research 
group cannot afford such a long process anyway.  

Over the years, many scientific visualization ideas and techniques have 
been introduced, but only a few have been actually deployed and routinely 
used by scientists. We argue the absence of user-centered design 
in the development process is the primary reason. Most of the scientific 
visualization designs are evaluated with case studies or a comparative 
performance study, without going all the way to deploying the 
new capability into scientists' workflow. Some of the techniques have 
been included in open source visualization software toolkits 
such as VisIt~\cite{VisIt} and ParaView~\cite{ParaView}, which find many users. 
However, for scientists like our collaborators who have been 
developing computational simulation of turbulent reacting flows 
with complex chemistry and pushing computing’s limits to achieve scientific 
discovery, a custom design of data analysis and visualization solution
is more desirable.
Over the course of two years, we have collaborated with domain scientists at 
a U.S. National Laboratory to develop visualization solutions for them. 
Our collaborators develop large-scale turbulent combustion simulations 
and run these simulations on some of the most powerful supercomputers 
in the world. The amount of data these simulations capable of outputting can 
be overwhelmingly large, presently from hundreds of terabytes to several petabytes. 
As the supercomputing power continues to grow, the scale of the data is expected
to be even greater and the traditional post hoc data analysis approach
is no longer feasible. Scientists have thus turned to in situ visualization solutions 
for possibly validating and analyzing their simulation results. 
In situ data processing must be coupled with the simulation, either tightly or loosely,
and executed in the same supercomputing environment, presenting many unique 
challenges to the design and implementation of the solutions. 
The in situ processing may be to compute visualizations, perform data reduction, 
or conduct feature extraction. We, the visualization researchers, must communicate with
the scientists to know not only the simulated phenomena but also the simulation code itself
to sufficient extent so that we can derive a viable design.  
Our collaborators, the scientists, need to understand the capabilities of state-of-the-art 
visualization techniques  to effectively participate in the design process.


The desire for usability suggests a user-centered design. 
Our collaborators' committed effort makes possible an iterative design process, deriving
a solution for them consisting of an in situ code computing a statistical summarization of select flow field regions  and a post hoc code allowing them to 
interactively review the simulation results in terms of the statistical summarization~\cite{Ye:2016}. 
This new capability enables our collaborators to have a greater control of what they consider important to focus on analyzing at the needed detail and fidelity, which effectively increases the outcome and value of their simulation-based study.

In this paper, we present the user-centered design process 
used to produce the in situ visual analysis technology 
for our collaborating scientists~\cite{Ye:2016}.
In this process, we have also refined the visual 
interface design to better support the post hoc analysis 
tasks. In summary, our work makes the 
following contributions:

\begin{itemize}
\setlength\itemsep{0pt}
    \item We practice and demonstrate a
    user-centered design for the development of 
    a scientific visualization solution based on 
    in situ computing 
    of regional probability distribution functions to support 
    the data analysis needs of large-scale combustion simulations. 
    
    \item We adapt a user-centered design process for working 
    with a particular group of domain experts to 
    develop a unique, usable data analysis solution and 
    report the necessary adjustments to the process 
    and their underlying rationales.

   
   \item We report the lessons learned through a design
    process over a two year long collaboration with combustion scientists, which should aid others seeking to develop usable solutions in a similar manner. 

\end{itemize}

\section{Related Work}


Making sure the proposed visualization techniques/systems are useful, has been a focus in the recent visualization works. To achieve this goal, many researchers have incorporated extensive evaluations. Lam et al.~\cite{Lam2012-rk} and Isenberg et al.~\cite{Isenberg2013-bm} performed comprehensive reviews of visualization literature and came up with a categorization of evaluations.
As indicated by their research, researchers in scientific visualization have been mostly using algorithmic performance and case studies as the evaluation methods. Even though they are both very important techniques, it is generally not enough to demonstrate the usability of a visualization system. With an increasing focus on evaluations, user experience has become an important aspect. In this context, Roger et al.~\cite{Rogers2011-lx} and Saket et al.~\cite{Saket2016-sc} suggested that visualization designs should meet both usability and user experience goals.



It is extremely important to adhere visualizations to the domain experts' needs and requirements.
As highlighted by Johnson~\cite{Johnson2004-io}, visualization experts need to be ``working side-by-side with end users to create better techniques and tools for solving challenging scientific problems'' as the top research problem of scientific visualization. Later in the year of 2007, Ma et al. organized the IEEE Visualization panel ``Meet the Scientists'' \cite{Ma2007-dg} to promote direct interactions between visualization experts and scientists. They believed ``this interaction is crucial for obtaining the understanding of what scientists really need to get out of their datasets and what visualization functionalities are missing in existing visualization software tools.''

The term user-centered design was first used by Donald Norman~\cite{Norman:1986,Norman:2002}, who described a design process that primarily focuses on specific needs of the user rather than less important factors, such as aesthetics. The topics of usability engineering and usability testing were heavily influenced by user-centered design as presented in multiple books~\cite{Nielsen1994-pr,Jeffrey1994-du,Dumas1999-ml}. Later in 2010, an ISO standard~\cite{Standardization2010-zw} was established on human-centered design processes for interactive systems, which identifies key activities as: understanding context of use, determining requirements, producing designs, and performing evaluations.

The information visualization community has been utilizing user-centered design processes and developing guidelines for design studies. Sedlmair et al.~\cite{Sedlmair2012-ck} presented a methodological framework for conducting design studies. They based the guidelines and pitfalls presented on the authors' combined experiences and an extensive literature review.
Lloyd and Dykes~\cite{Lloyd2011-yz} provided invaluable recommendations on how to incorporate user-centered design in developing geo-visualizations by presenting the experiences they gained from multiple cases in a long-term study.
Similar works have also been presented to share the experiences of adopting a user-centered design process~\cite{Slingsby2012-gs,Wood2014-gh,Espinosa_undated-sa}. On a finer granularity, Vosough et al. in \cite{Vosough2017-ss} aimed to specifically analyze how to better establish requirements in real-world visualization projects. These works have done an effective job of promoting user-centered design, as more and more works in information visualization~\cite{Komlodi2005-th,Dykes2007-co,Dong2008-ae,Goodwin2013-mt,McKenna2015-vc} have adopted the idea.


However, user-centered design is less formalized in the field of scientific visualization, other than some works~\cite{Kerren2007-ld,Sedlmair2012-ck,Klein2012-yq} toward data visualization in general, and some call to actions~\cite{Johnson2004-io,Ma2007-dg} that suggest visualization experts in scientific visualization should directly interact more with domain experts. Some aspects of user-centered design have been naturally adopted by scientific visualization, such as the concept of working closely with the scientists and performing case studies. Other aspects have not been fully utilized. In particular, usability studies are a major component of user-centered design and have been successfully introduced to many information visualization works~\cite{Wu2009-js,Bateman2010-ai,Tory2014-km} among many others, but have yet to make a major appearance in the field of scientific visualization.

User-centered design is potentially even more important in the field of in situ visualization. The evaluation of the effectiveness of an in situ visualization technique has traditionally been focused on algorithmic performance, as researchers are developing new performance models specifically for in situ visualization and analysis~\cite{Ayachit:2016:PAD:3014904.3015010, Larsen:2016:PMS:3014904.3014936}.
A complete solution of in situ visualization often involves two parts.
Firstly, in situ routines need to be injected into the scientific simulations in a user-friendly way.
There have been some works on improving the situation by introducing in situ infrastructures~\cite{ayachit2015, whitlock2011, lofstead2009, Vishwanath:2011:TDM:2063384.2063409}. Secondly, a visualization component is often required to process and analyze the in situ generated results.


Our solution to generate regional probability distributions functions (PDFs) in~situ~\cite{Ye:2016} can be considered as a data preprocessing technique~\cite{kim2011, su2015, zheng2013}.
The use of histograms has also been studied in various areas. Novotny et al. \cite{novotny2006} utilized histograms to generate parallel coordinates in real time for interactive exploration of large datasets. Thompson et al. \cite{thompson2011} used hixels (1D histograms) to represent either a collection of values in a block of data or collections of values at a location in ensemble data. They demonstrated that topological analysis and uncertainty visualization can be performed with hixels. Neuroth et al. \cite{neuroth2015} generated spatially organized velocity histograms both on-the-fly and in situ for interactive visualization and exploration of the underlying data.
Our approach uses in situ regional multidimensional PDFs to provide post hoc analysis at a high granular level.


\section{Design Study Overview}

There are many proposed user-centered design processes. Traditional design processes~\cite{Slocum2003-fa,Robinson2005-di} begin with domain analysis to identify problems and solutions before moving onto prototyping. Roth et al.~\cite{Roth2010-hj} and Koh et al.~\cite{Koh2011-vr} provide alternative processes that use quick prototyping methods (e.g., paper prototyping) before domain analysis to better demonstrate the potentials of visualization techniques.
In our case, we believe domain analysis is necessary to bridge the knowledge gap between the domain scientists and us. Also, we need to experiment with the implementation of the in situ routines before we can come up with an interface to communicate with the simulation.
As a result, we opt to adapt a more traditional user-centered design process.

\begin{figure}[ht]
 \centering
 \includegraphics[width=0.95\columnwidth]{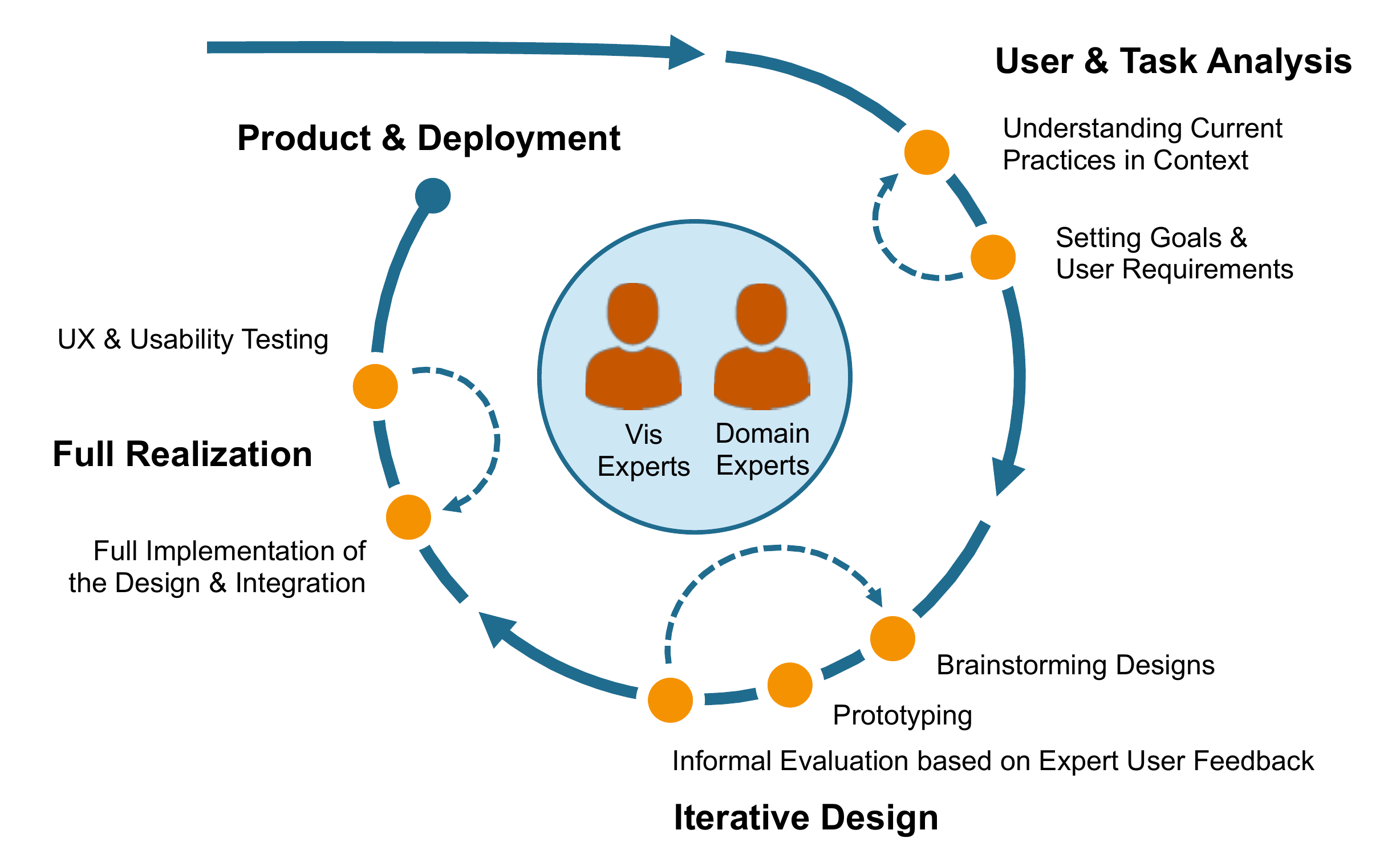}
 \caption{A visual depiction of our user-centered design process. Both the visualization expert and domain expert are heavily involved in the many iterative loops that occur throughout the overall process.}
 \label{fig:ucd:process}
\end{figure}

For the remainder of this manuscript we describe an extensive collaboration with combustion researchers at a U.S. national laboratories, over which we adapted a user-centered design process in order to develop effective and usable visualization solutions.
The process we applied, as depicted in Figure~\ref{fig:ucd:process}, consists of three stages.
At the beginning, the \textit{User and Task Analysis} stage focuses on understanding the context of the domain problem, setting a set of goals, and identifying user requirements, 
as reported in detail 
how we bridge the knowledge gap before setting goals in Section~\ref{sec:domain_analysis}.
Then, the collaboration enters the \textit{Iterative Design} stage, in which we iterate through sketching a design, prototyping the design, and frequently soliciting feedback from domain scientists on particular aspects of the design, as discussed in Section~\ref{sec:iterative_development_stage}. 
Note that in this stage, it is possible for the initial goals and requirements to change based on new information learned throughout the prototyping process. We experienced such a design shift in this stage, which is described in Section~\ref{sec:ucd_shift}.
The final stage is the \textit{Full Realization} stage, as discussed in  Section~\ref{sec:iterative_refinement_stage}, which consists of 
detailing the design and its full implementation, followed by two types of usability testing: \textit{User Experience Testing} (Section~\ref{sec:user_experience_testing}) and \textit{Domain Expert Usability Testing} (Section~\ref{sec:domain_expert_usability_testing}). 
Note that in the rest of the paper, we use {\em domain experts, domain scientists, scientists}, and {\em users} interchangeably. 

\section{User \& Task Analysis}
\label{sec:domain_analysis}



The objective of the first stage of the design process is to obtain a full understanding of the users and tasks. However, this is not an easy job because of the knowledge gap between the visualization experts and the domain experts. On one hand, domain specific knowledge is important for visualization experts to truly understand the problems raised by the domain experts. On the other hand, visualization knowledge allows domain experts to identify better problems that are suitable for visualization.
After there is a sufficient level of mutual understanding, the next step is to identify problems and establish requirements by developing \textit{Goals and User/Task Requirements}. The problems usually arise naturally as the mutual understanding develops. They can be identified from existing time-consuming tasks, lack of capabilities from existing tools, etc. Based on the problems and requirements, an initial design is developed. Such a design is not intended to be set in stone and should be reviewed and refined in the next stages when appropriate.

Many user-centered design methods can be used for such domain-specific analysis. Koh et al.~\cite{Koh2011-vr} and Slingsby and Dykes~\cite{Slingsby2012-gs} operate through
a visualization awareness workshop 
to educate the domain experts about general visualization concepts. Specifically, the visualization experts present a range of data visualization techniques to the domain experts in the workshop. Slingsby and Dykes~\cite{Slingsby2012-gs} also organize another
workshop, in which the domain experts demonstrate their 
current practices. In our design process, we choose to conduct
similar working sessions with the scientists  
to help close the knowledge gap.
Lloyd and Dykes~\cite{Lloyd2011-yz} recommend to establish a master-apprentice relationship instead of the client-consultant style at this stage as to gain trust from the domain experts. They do interviews to gather contextual information and use methods like card sorting and calculating word frequencies to process the gathered information.
In our approach, we take this a step further and treat 
the domain experts as team members, encouraging them to contribute directly to the actual development of a technique or strategy.
From our experience, we believe it is less about following a particular set-in-stone procedure for domain-specific analysis. 
Instead, we should choose methods/techniques that suit 
all parties to facilitate the communication,
establish a strong relationship with the expert users, 
and develop an actionable plan.

\subsection{Our Process}

We began our user \& task analysis stage with a series of working sessions, similar to the visualization workshops described by Slingsby and Dykes~\cite{Slingsby2012-gs}, in which the visualization researchers could formally present relevant state-of-the-art visualization techniques. 
In our case, the choice of techniques were initially based on 
a limited understanding of the domain experts' workflow from prior collaborations. We chose to include topics from in situ visualization~\cite{Bauer2016-kx,Kress2016-iv}, scientific storytelling~\cite{Ma2012-zh,Hoffman2014-zw}, statistical analysis of volumetric data~\cite{Chaudhuri2014-os,Thompson2011-gl}, and particle trajectory analysis~\cite{Salzbrunn2008-qe,Burger2008-nt}.
Our primary goal was to educate the domain experts about advanced visualization concepts, inspire them to identify both opportunities and challenges in their workflow that are suitable for visualization, and discover collective interests among both parties. Because a mutual understanding between both teams is so fundamental to this step, we still held this session despite the fact that we had collaborated with some of these scientists in the past.

The outcome from our initial working session was ideal 
in the sense that the domain experts displayed excitement upon learning about the different visualization techniques available, enabling them to immediately identify how these techniques can help in their work. We then continued our collaboration with another session in which the domain experts  demonstrated the different problems they are facing and their current practices. 
The combustion scientists demonstrated their state-of-the-art simulation and their unique way of using particles to help analyze the 3D field data. Much of this session was very demonstrative in nature and took place in the scientists' common work environment. The primary goal was for the visualization researchers to observe typical analysis tasks to obtain a complete understanding of the scientists' workflow.
We encouraged domain experts to come up with ideas without thinking about how practical the ideas are. Without thinking what visualization techniques can solve their problems, they came up with ideas that tailor more to their actual problems, which helped us understand what they really desired. This practice ended up generating ideas that are effective and more practical.   

Once the knowledge gap between us became narrowed, a set of follow-up meetings were conducted to discuss potential collaboration topics with a goal of identifying those with the largest impact. These meetings are more informal than the working sessions and include an open discussion format with the aim of brainstorming as many ideas as possible. In our case, the combustion researchers expressed immediate interest in applying many of the visualization techniques to their workflow while we proposed new techniques based on our, now improved, understanding of the scientists workflow. The discussions then continued to determine the challenges involved in applying these techniques to a simulation at scale and what new methods might need to be developed to meet our goals.

After several in-person and remote meetings via teleconference, we were able to identify key topic areas including in situ visualization, data distribution analysis, and particle trajectory analysis which would have the largest impact on the scientists' workflow. For each topic area, we then began to narrow down a set of potential solutions for improving the domain experts' scientific process and evaluated their feasibility. 
Follow-up communications then extended on these directions to generate concrete requirements and actionable plans. 





\subsection{Common Practices in Context}
\label{current_practices}

\begin{figure}[ht]
	\centering
	\begin{subfigure}[b]{0.24\columnwidth}
		\includegraphics[width=\textwidth]{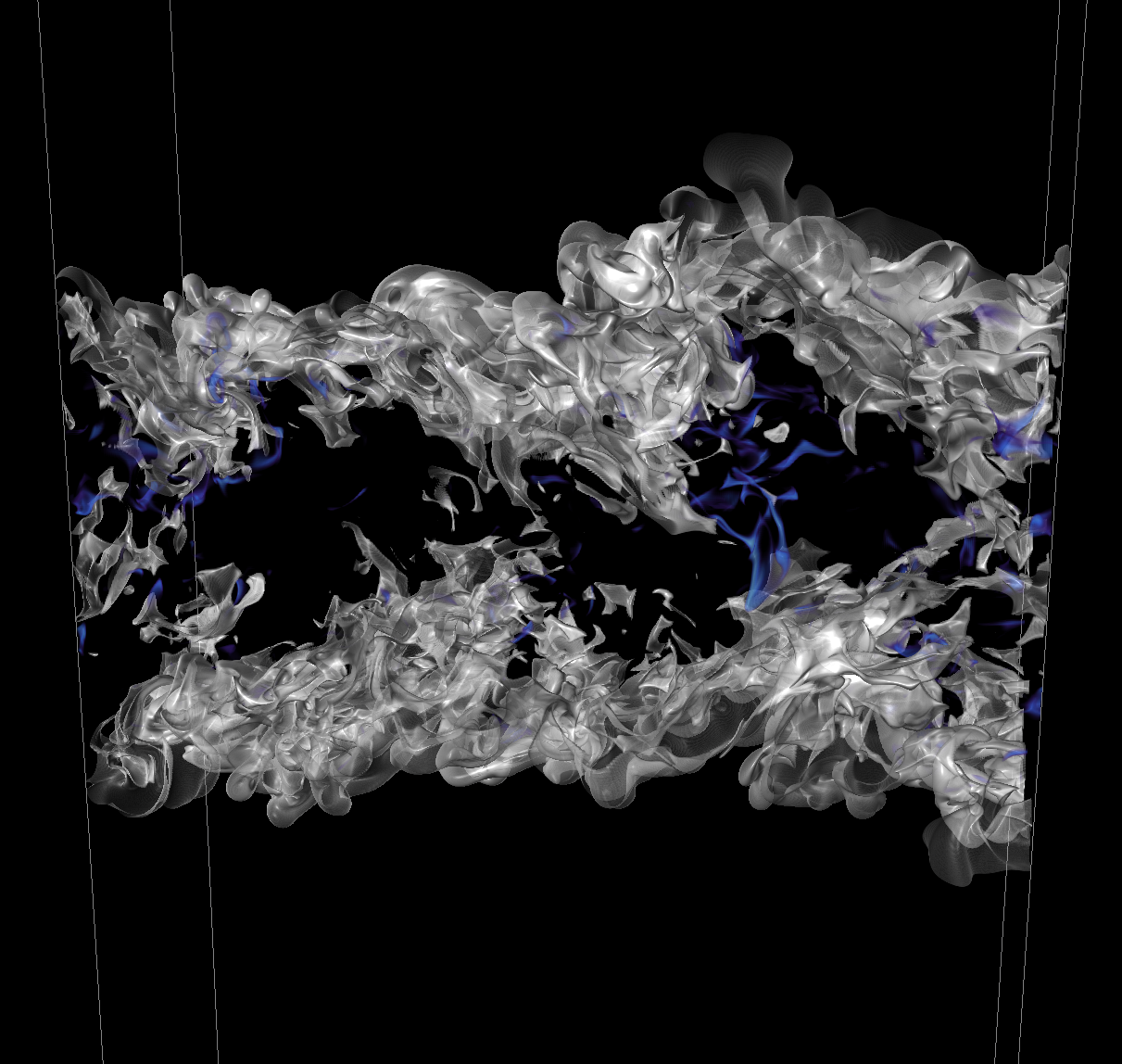}
		\caption{Before}
		\label{fig:s3d:begin}
	\end{subfigure}%
	\hspace{\fill}
	\begin{subfigure}[b]{0.24\columnwidth}
		\includegraphics[width=\textwidth]{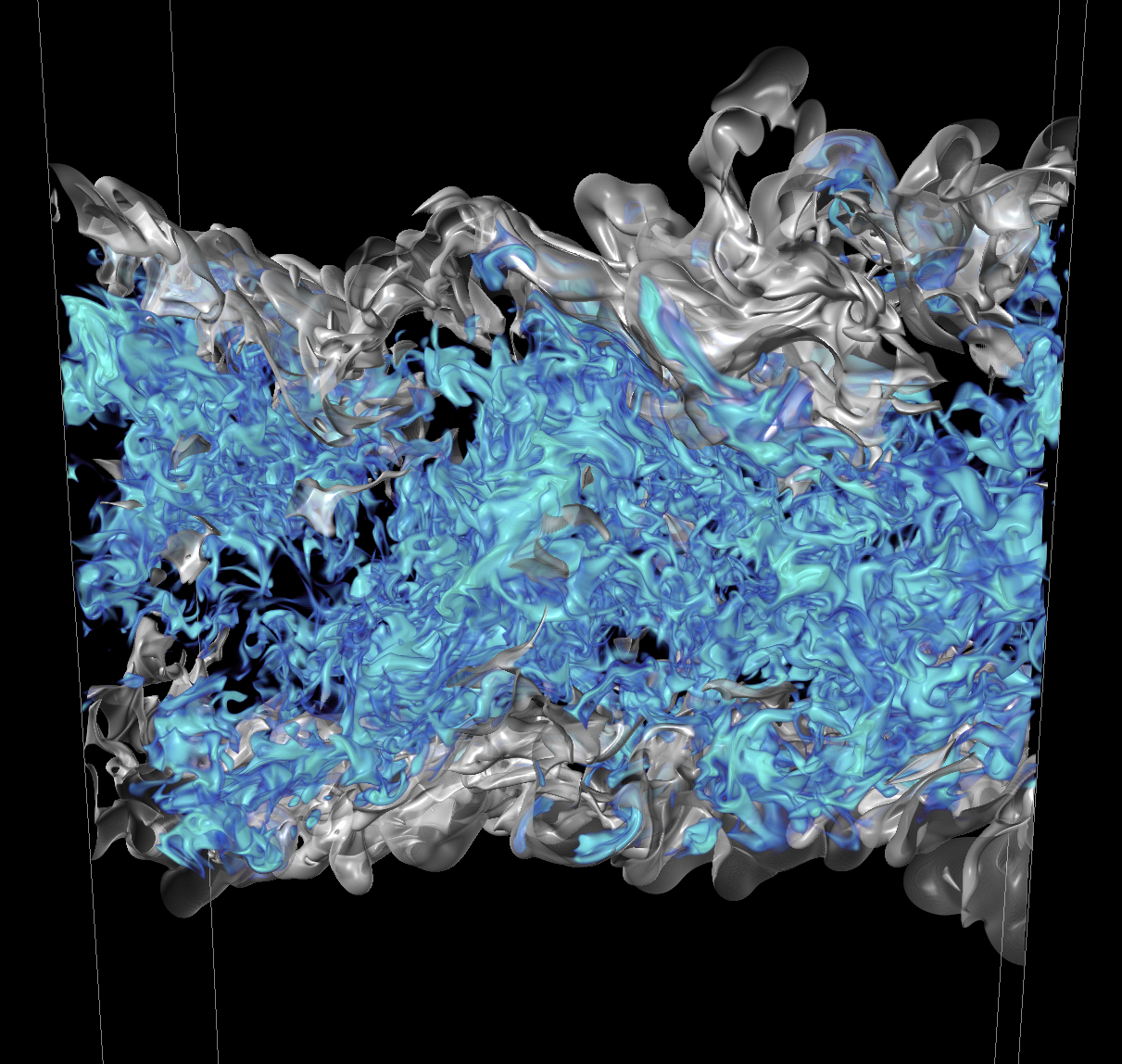}
		\caption{Stage One}
		\label{fig:s3d:low_temp}
	\end{subfigure}%
	\hspace{\fill}
	\begin{subfigure}[b]{0.24\columnwidth}
		\includegraphics[width=\textwidth]{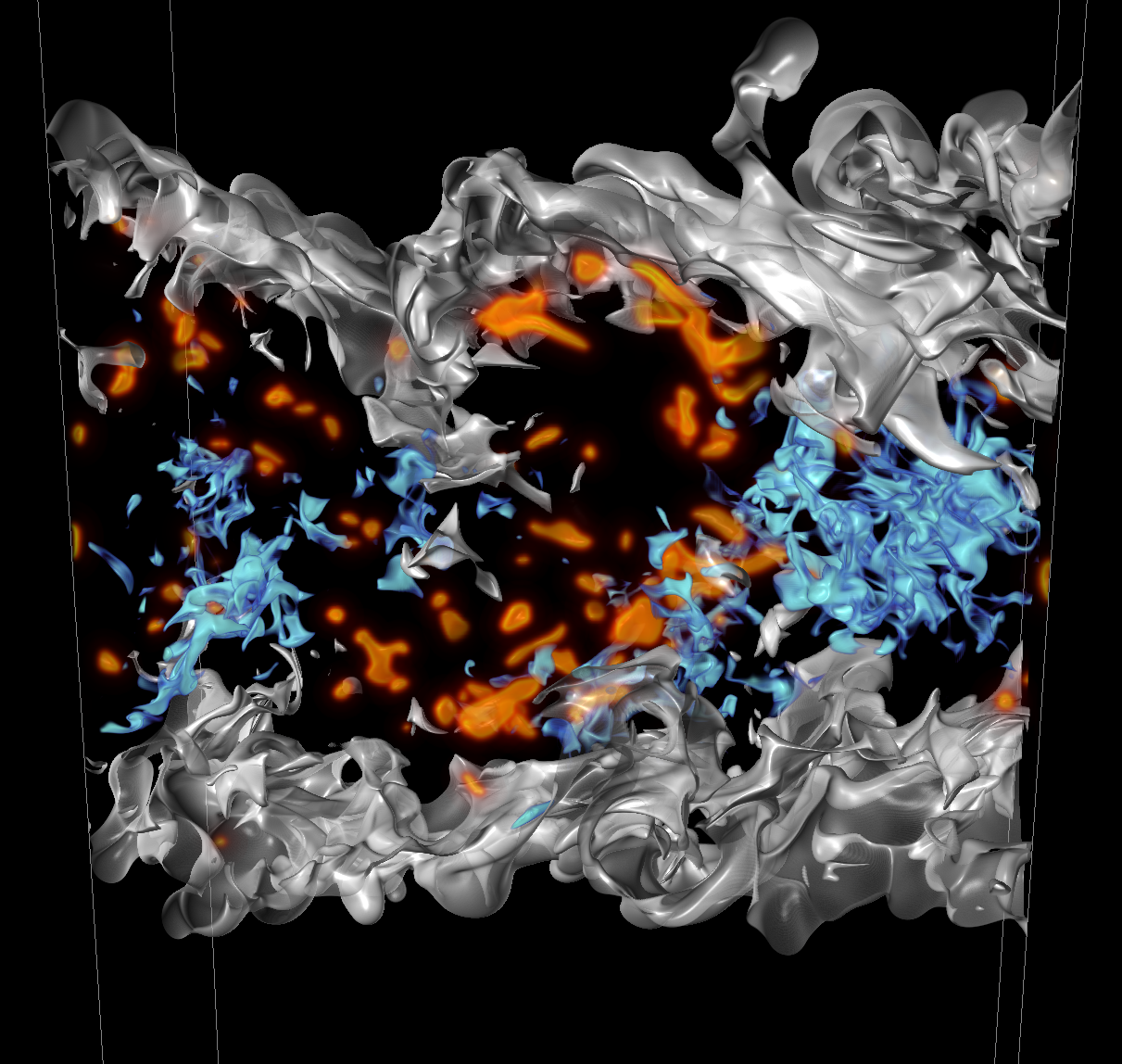}
		\caption{In-between}
		\label{fig:s3d:switch}
	\end{subfigure}%
	\hspace{\fill}
	\begin{subfigure}[b]{0.24\columnwidth}
		\includegraphics[width=\textwidth]{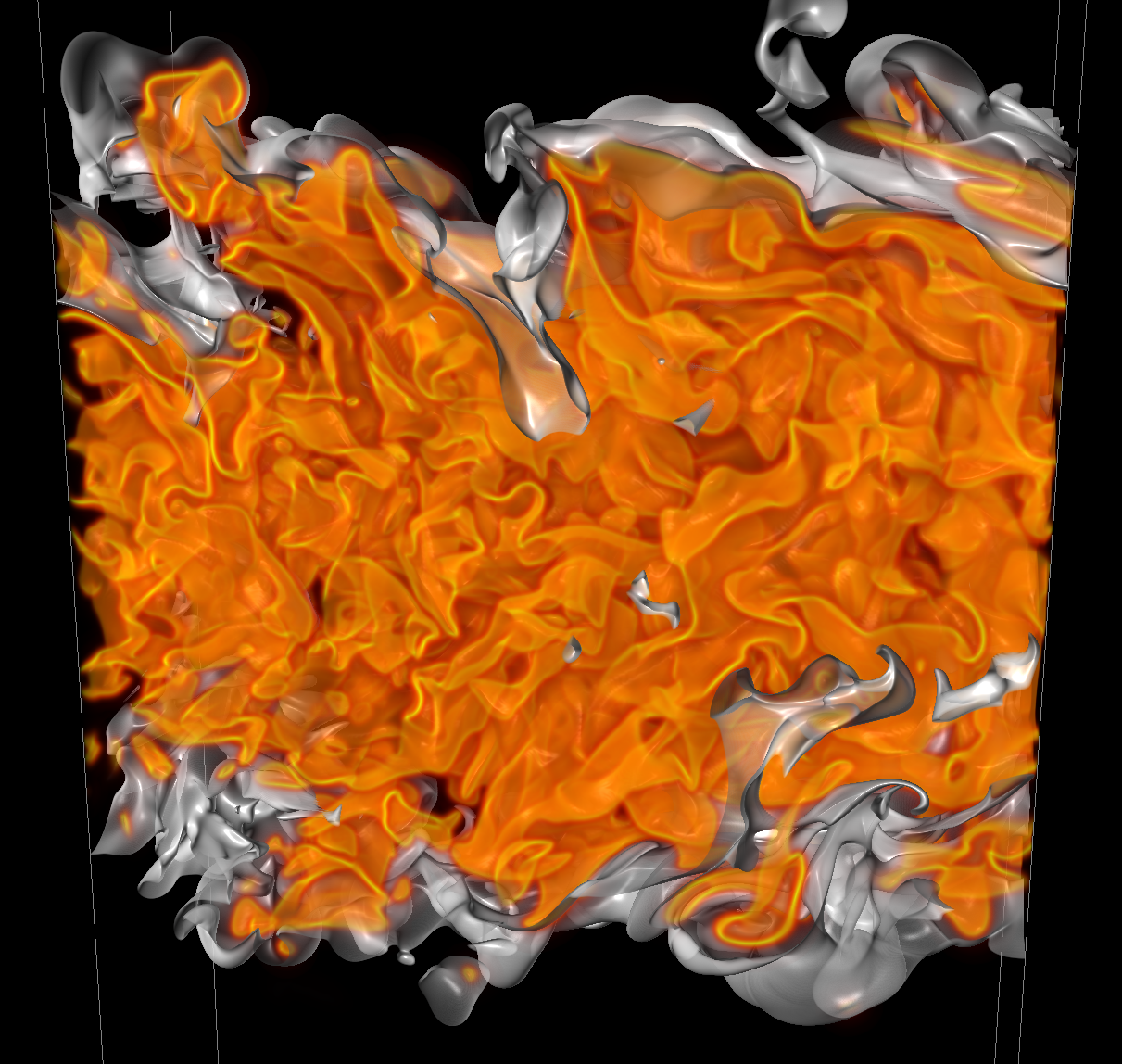}
		\caption{Stage Two}
		\label{fig:s3d:high_temp}
	\end{subfigure}
	\caption{A large-scale simulation depicting two-stage auto-ignition of n-dodecane enabled by S3D. The blue features indicate low temperature ignition by highlighting keto-hydroperoxide. The orange features show a constant temperature value to indicate high temperature ignition.}
	\label{fig:s3d}
\end{figure}

Through extensive discussion with the domain scientists, 
we were able to understand their common practices. 
Their primary focus is the study of combustion systems using S3D~\cite{Chen2009-ir}, a petascale combustion simulation code developed by researchers at Sandia National Laboratories. 
The domain experts utilize S3D to perform vastly different simulation runs for studying fundamental turbulence-chemistry interactions in combustion processes, for the design of efficient and clean engines reliably burning a diverse range of fuels. 
Visualization of a typical simulation run is shown in Figure~\ref{fig:s3d}.
Often times, the scientists would modify the S3D simulation code, which is primarily written in Fortran, according to a particular 
problem of interest. The simulation code mostly uses non-uniform rectilinear grids, but also uses a new multiblock configuration for modeling more complex domain shapes.

With the advancement of supercomputing, today S3D is able to output  massive amounts of data, near the scale of petabytes. As a result, I/O became a major bottleneck in the domain experts' workflow causing them to look towards in situ visualization as a potential remedy. The I/O routines of S3D were mostly using Fortran I/O and MPI I/O. The domain experts attempted to integrate an early version of ADIOS~\cite{lofstead2009}, an existing, general-purpose in situ infrastructure, but the effort was later discontinued due to unsatisfactory usability. 
To deal with the overwhelming amount of data, the domain experts had no choice but output fewer time steps.
However, the reduced temporal resolution limited the fidelity of their analyses.
In some simulations tracer particles are advected to study the Lagrangian statistics.
The tracer particle data is also output periodically to perform analyses.
The amount of particles was orders of magnitude smaller than the number of grid points. To help put things into perspective, a particular simulation run generated $564GB$ of field data and $50GB$ of particle data per time step.

The primary analysis of the S3D data was done in post processing. 
For statistical analysis, the scientists utilized MATLAB~\cite{MathWorks1992-ya}. As for visualization tools, 
they used VisIt~\cite{Childs2012-yd} and occasionally ParaView~\cite{Ayachit2015-zo}. They were already familiar with multiple visualization techniques, such as 3D contours, 3D distance functions, and direct volume rendering. Since even one time step of the simulation data was too large for a local workstation to handle, they used TECPLOT ASCII files~\cite{Bellevue2003-xs} to only visualize slices of the simulated 3D flow field. 
Though it was able to yield results, it was far from 
an effective solution for routine work. 

The combustion simulation team provided us both the source code of S3D with a simple case specification and existing datasets. This would facilitate our development of prototype modules that could be directly coupled the simulation code. They also explained the background scientific phenomena of the datasets so that we had a better understanding about what we were looking at.

\subsection{Goals and User/Task Requirements}
\label{goals_and_plan}

Based on the results of user and task analysis, we identified three technological directions to pursue: in situ visualization, distribution analysis, and particle trajectory analysis. 
We then established a primary goal of enabling efficient particle extraction in post hoc processing based on the high resolution field data. This was then further divided into three requirements. First, particle filtering and extraction must be done without iterating through all the particles. Second, there must be a responsive interactive user interface for domain experts to experiment with different particle filtering criteria. Finally, while the field data can best facilitate the particle filtering, it is too big to output entirely. We need a compact representation of the field data that  still preserves its high temporal and spatial fidelity.  
Then, the post hoc visual analysis of the particle data can be carried out with an average desktop workstation.

\begin{figure}[ht]
	\centering
	\begin{subfigure}[b]{0.49\columnwidth}
		\includegraphics[width=\textwidth]{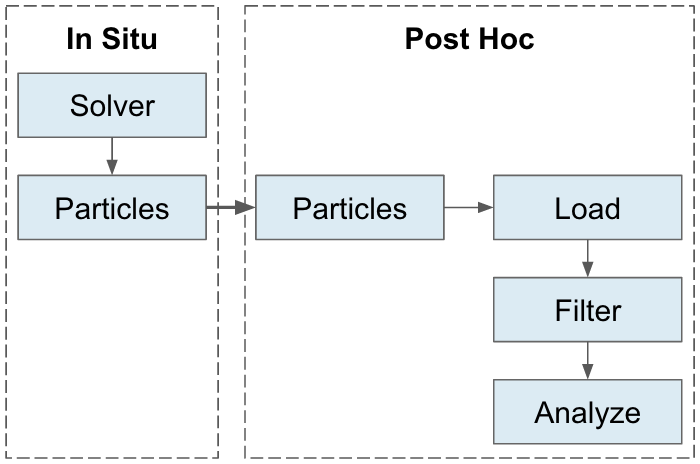}
		\caption{Original Workflow}
		\label{fig:workflow:old}
	\end{subfigure}
	\begin{subfigure}[b]{0.49\columnwidth}
		\includegraphics[width=\textwidth]{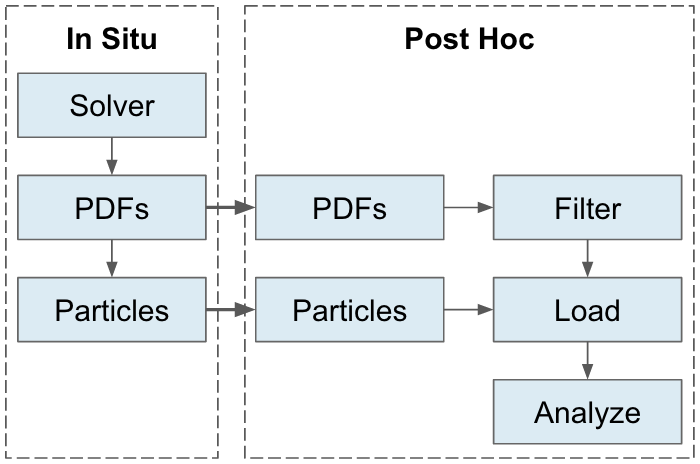}
		\caption{Proposed Workflow}
		\label{fig:workflow:new}
	\end{subfigure}
	\caption{The scientists' workflow to select and analyze simulation outputs. In the original workflow, the entire set of particles is filtered by iterating through each particle, which is time consuming. The proposed workflow is to apply filters to  probability distribution functions (PDFs) generated in situ and only load the particles that match the filtering criteria, thus allowing interactive particle selections~\cite{Ye:2016}.}
	\label{fig:workflow}
\end{figure}

Meeting the three requirements suggests a workflow, as shown in Figure~\ref{fig:workflow:new}, utilizing in situ generated probability distribution functions (PDFs) to aid post hoc particle selection and analysis.
PDFs are commonly used to provide an overview of data. In our scenario, PDFs provide a great intermediate representation of the
original field data because they drastically reduce the data size and can capture a great amount of statistics for domain experts to make sense of the simulations. Furthermore, since the simulation data are spatially organized, we can subdivide simulation volumes into regions and generate PDFs per region, thus providing even more statistics for the domain experts to explore.
With such regional PDFs, the domain experts are able to identify combustion simulation events in post processing, and it is even possible to automatically detect features of interest in situ.
In addition, particles are sorted according to the same spatial organization scheme of the PDFs before being written to high performance storage. In post processing, instead of filtering the particles directly (a time consuming operation), the scientists can interactively apply filters to the PDFs in an exploratory fashion. Particles are then loaded from disk according to the selected PDFs. Since both the particles and the PDFs use the same spatial organization scheme, our post hoc visualization tool is able to quickly pinpoint the memory locations of the particles on disk without scanning through the entire dataset. 
An important design goal was to make all interactions on PDFs to be responsive for the scientists to verify their ideas quickly. The particle loading operation, which is based on the filtered PDFs, primarily depends on the amount of particles that meet the query and need to be loaded into memory. However, since the PDF selections already subsample the domain, this amount tends to be small enough to achieve a usable level of interactivity.

\section{Iterative Design}
\label{sec:iterative_development_stage}


There are two main steps involved in this iterative process. The first step is \textit{Brainstorming Designs}. In many cases, there are multiple design decisions that need to be made and it is not always clear which will work best. This step lays out potential options for each of these decisions. In contrast to the user analysis stage, this step can reflect on the knowledge gained from the iterative process to make better decisions on both project details and directions. 

The second step is \textit{Prototyping}, which involves simple and quick implementations to test out each of the options and determine which will work most effectively. This continues in an iterative fashion until all design decisions are made and a complete working prototype is ready for testing. It is possible to make major adjustments to the direction of the project at this stage. This is manly due to inspirations provided by the prototypes, as newer ideas and more suitable solutions can be discovered. In this step, both the domain experts and visualization experts are heavily involved in making each of the design decisions. Formative usability studies are commonly performed in this step.

Wireframes and mock-ups can be used in early iterations to convey the basic ideas and designs. The point of these prototypes is that they can be developed quickly thus allowing visualization experts to quickly test their essential ideas. The quick turnaround time of these prototypes is important as ideas can be abandoned at this stage. For data visualization, prototypes can be `hacked together' using existing software packages, such as ParaView~\cite{ayachit2015}, VisIt~\cite{Childs2012-yd}, and D3.js~\cite{Bostock2011-hl}. To evaluate the user interactions, \textit{chauffeured} prototyping~\cite{Lloyd2011-yz} and \textit{wizard of oz} prototyping~\cite{Dahlback1993-sj} can be commonly used. Such early prototypes provide a common ground for both the domain experts and visualization experts to develop upon.

However, limitations involved in developing in situ visualizations often require more sophisticated prototypes.
In our case, we need to get familiar with the simulation code base and experiment with injecting code into the simulation before we can propose a prototype to the domain experts. By the time we are familiar enough with the simulation code base, we already have a working prototype of the in situ routines.
As a result, digital prototypes are used early on in the iterative design process. These digital prototypes are able to provide a much more realistic user experience to the domain experts. Slingsby and Dykes~\cite{Slingsby2012-gs} and Lloyd and Dykes~\cite{Lloyd2011-yz} recommend to use the domain experts' data at this stage because it would be easier for them to assess the unfamiliar visualization techniques.


\subsection{Brainstorming Designs and Prototyping}
\label{sec:prototyping}

Distance constraints limited our physical meetings with the domain experts to every two to three weeks, but we remained in constant remote contact to brainstorm and make detailed design decisions. One example was to decide on how to organize the PDFs in the post hoc visualization tool. The initial decision was to have a list-like view to display all the PDFs with extracted ones placed in the front of the list. After further iterations with the domain experts, we came to the idea to display the PDFs according to their spatial relationship, which not only gave structure to the PDFs, but also allowed users to observe spatial patterns in the simulation volume. These brainstorming sessions also helped us discover usability problems and continuously steer the project direction.

Our prototype was naturally separated into an in situ library and a post hoc visualization tool. Since an in situ infrastructure was not present in S3D, the in situ library had to be directly injected into the simulation. Thus, there were limitations in what ways our tools could communicate with the simulation. We ended up implementing the in situ library using C while exposing a Fortran interface to communicate with S3D. During the development of the in situ library, we relied heavily on the domain experts' knowledge with the S3D code base. Furthermore, they directly contributed an adapter for the Fortran interface. Since the in situ library was to be run alongside the simulation in a supercomputing environment, it had to be robust and scalable.  We wrote a test suite for the library to ensure it was functioning as expected. Even though the in situ library was designed to be run in situ, we soon realized that it is important to also have a separate post hoc application that could generate PDFs from existing datasets. This is because it would cost a great deal of resources to redo prior simulation runs.
Another important design consideration is how to specify PDF configurations, such as value range and bin width, as a good configuration is important to capture the correct data distribution. At the beginning, we decided to keep it simple and manually specify a global PDF configuration for all regions and for all time steps of a simulation run. However, we later found that it was impossible to generate meaningful results with this approach because simulation data are continuous in nature and change drastically across the simulation volume and across time steps. As a result, we implemented multiple strategies (Sturges' formula~\cite{Sturges1926-uy}, Scott's normal reference rule~\cite{Scott1979-od}, Freedman-Diaconis' choice~\cite{Freedman1981-tb}, etc.) to automatically calculate the value range and bin width according to the underlying data distributions.
Due to the above reasons, the iterations on the in situ library took longer than we expected.

\begin{figure*}[t]
    \centering
    \begin{subfigure}[t]{\columnwidth}
        \centering
        \includegraphics[width=\columnwidth]{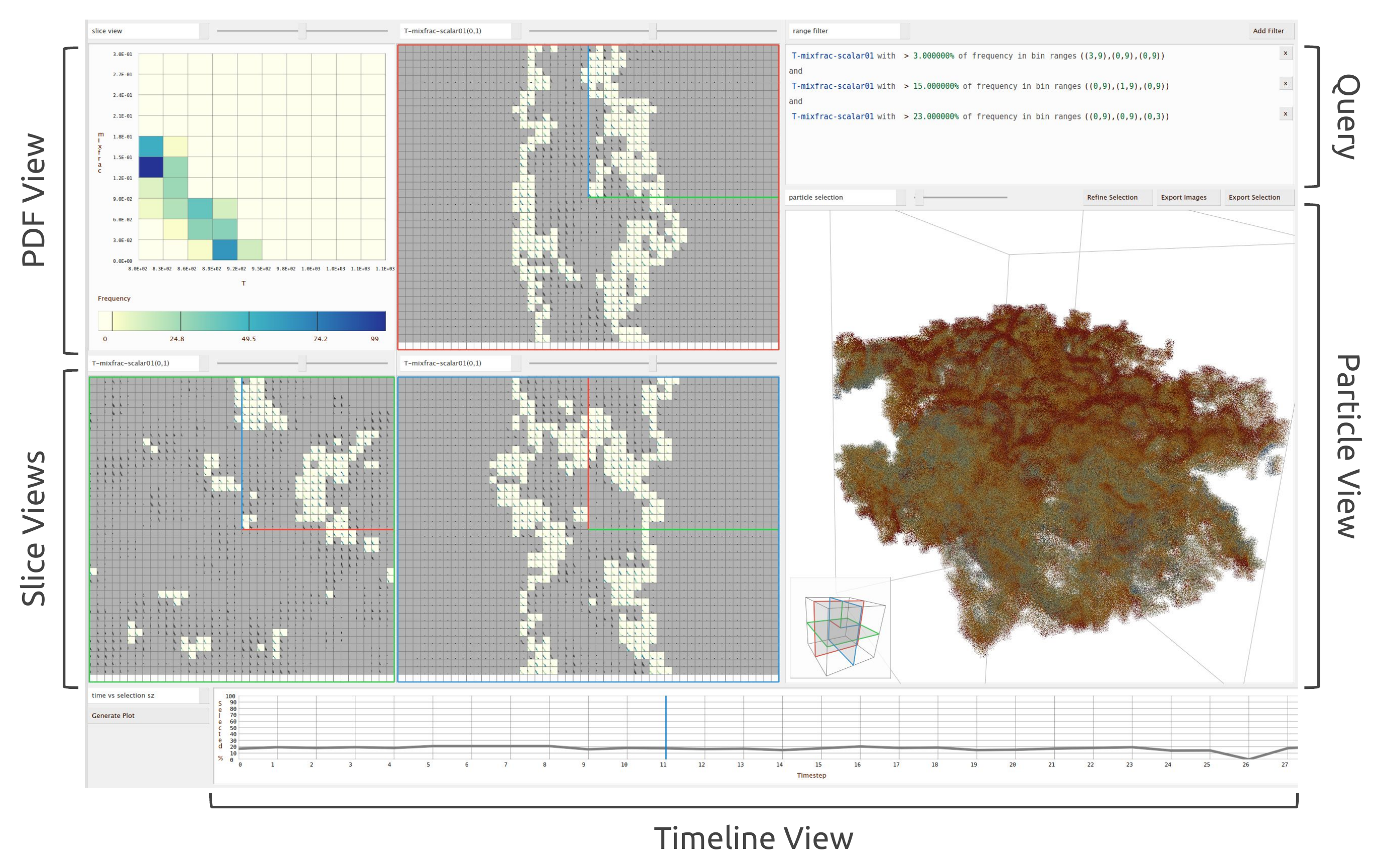}
        \caption{\textbf{Original Design:} 
        The three \textbf{slice views} show the PDFs on three orthogonal cutting planes in the volume. The \textbf{timeline view} enables time step selection as well as displaying the number of extracted PDFs per time step. The \textbf{query view} allows users to add and combine different filtering criteria for particle extraction. The \textbf{particle view} displays a 3D rendering of the extracted particles. Due to the large amount of visual elements shown and the limited screen estate, it is hard for users to distill insights, e.g. the PDFs in the \textbf{slice views} are too small to observe.}
        \label{fig:3d_vs_2d:3d}
    \end{subfigure}
    ~
    \begin{subfigure}[t]{\columnwidth}
        \centering
        \includegraphics[width=\columnwidth]{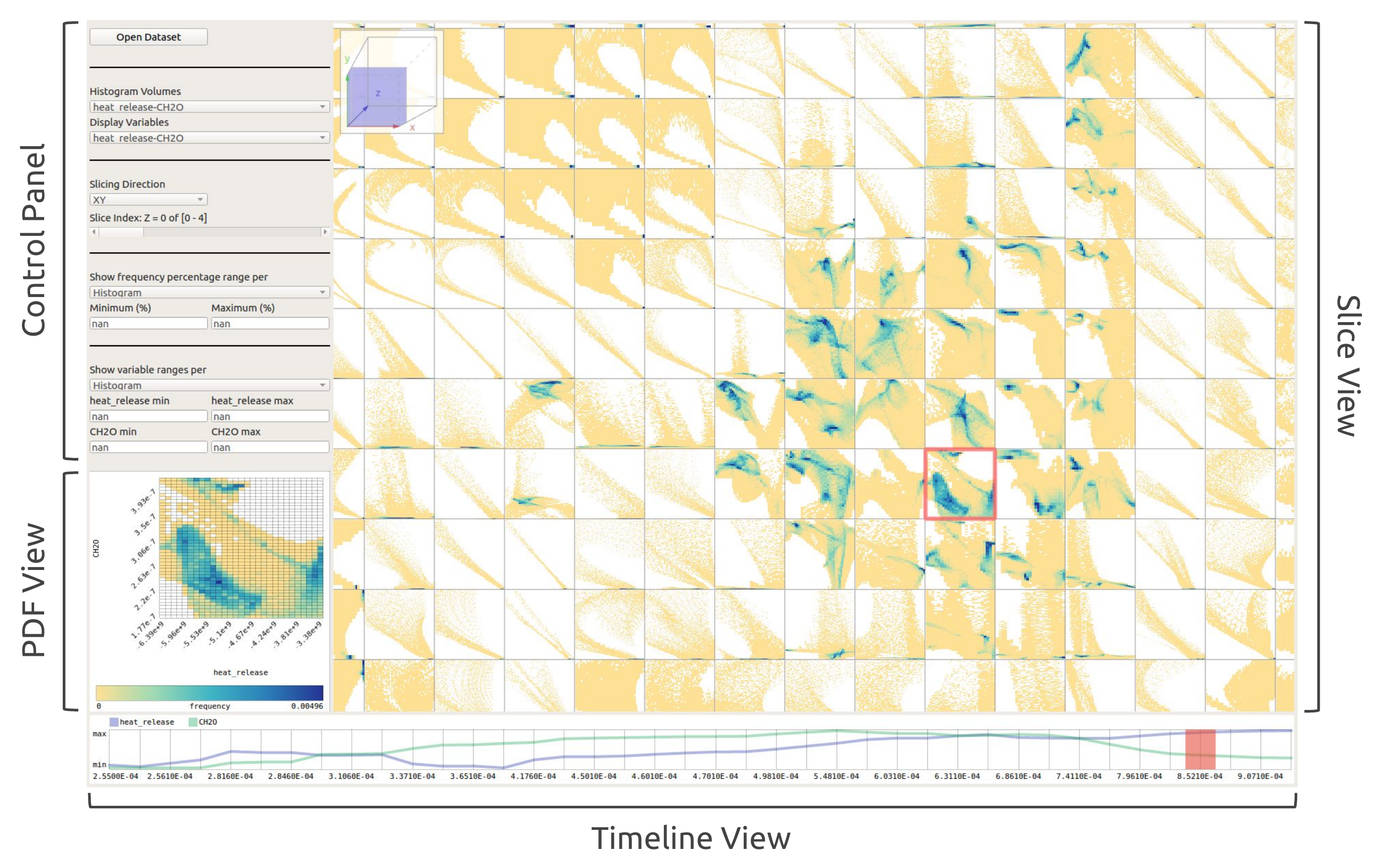}
        \caption{\textbf{Simplified Design:}  
        The \textbf{slice view} displays either 1D or 2D PDFs that are on an orthogonal cutting plane of the volume, and enables selection of PDFs as indicated by the red square.
        Zooming and panning are enabled for users to navigate the \textbf{slice view}. The \textbf{timeline view} enables time step selection as well as displaying the averages of variables per time step. The \textbf{control panel} hosts UI elements that are responsible for functionalities including data selection, PDF slice orientation, frequency range selection, and variable ranges selection. The fewer visual elements shown put more information in focus in this redesigned user interface.}
        \label{fig:3d_vs_2d:2d}
    \end{subfigure}
    \caption{Comparison between the user interface of the original design and the current simplified design. The simplicity of the current design makes the visualization tool easier to use, thus requiring less training. 
    Both designs use a \textbf{PDF view} for analyzing individually selected PDF.}
    \label{fig:3d_vs_2d}
\end{figure*}

The iterative process went faster with the post hoc visualization tool. In fact, a prototype of the post hoc tool was able to be implemented before the in situ library for the purpose of defining a data format for the generated PDFs. Since modifying the data format meant modifying the in situ library, we always tested the new data format with synthetic data to ensure it matched our needs. The frequent contact with the domain experts ensured the functionalities we implemented adhered to their requirements. The resulting interface is shown in Figure~\ref{fig:3d_vs_2d:3d}. The prototypes we implemented were more sophisticated mainly due to the requirement of handling large datasets. However, with the expectation that plans and requirements could change, we made sure to keep all aspects of the code flexible so user facing components could be easily rewritten and low level utility code could be reused.

When a new prototype was ready, we held a brainstorming session with the domain experts to gain feedback and discuss future prototype implementations. Feedback was generally gathered by applying heuristic evaluation and cognitive walkthrough. Feedback inspired new design decisions, and design decisions were reflected in new prototypes, thus completing the iterative loop from brainstorming to prototyping.

\subsection{The Pivot in Design}
\label{sec:ucd_shift}


As the prototypes became more mature, it became apparent that we need to refocus the functionalities of the visualization tool. As supporting features to enable efficient particle filtering, the PDF exploration and analysis features were gaining more and more excitement. With the prototypes, the domain experts gradually became more invested in the potential of the PDF features. As a result and per the scientists' requests, we applied the prototypes to other non-particle datasets they had. Through such experiments, we found that the statistical analyses provided by the PDFs are more flexible and beneficial than we originally thought.
After further communication with the domain experts about PDF usages in scientific data analysis, we found that PDFs are commonly used to understand the data distributions to identify plausible regimes in the observed phenomena and compute the relevant statistics to characterize the overall behavior. As simulation results are multivariate in nature, occurrences of significant events in the phenomena are associated with changes in the values of multiple variables. Analysis in such situation is often performed using joint statistics such as joint and conditional PDFs.
Thus, with further discussion with the domain experts, we eventually came to the conclusion that expanding the PDF analysis capability at the cost of dropping the support for particle analysis is the most effective direction forward. We also agreed that the particle analysis feature could be a future research topic.

With the new direction in place, we took the time to reflect on the aspect of user experience. A detailed comparison between the original and redesigned interfaces is shown in Figure~\ref{fig:3d_vs_2d}. The original user interface design packed a lot of information (e.g., the three slice views) and advanced features (e.g., particle filtering), thus extensive training was required to effectively use the tool. By reducing the number of visual elements shown to the users at a time, the user interface becomes easier to understand. For example, instead of showing all three slice views, the interface displays only one larger slice, thus allowing the users to better observe and explore it. On the other hand, by simplifying the controls, we made the tool easier to use yet more flexible. For example, instead of providing a sophisticated particle filtering language, we opted to provide simple text boxes for specifying the displayed variable ranges of the PDFs. By doing so, the tool loses the filtering feature but the users can better observe different patterns and trends in the data, which is more helpful according to the domain experts.

This pivot was the result of not correctly estimating the impact of each of the proposed features, but we learned that there is no way to avoid all mistakes during a collaboration with a knowledge gap. At the same time, we should embrace such pivotal moments as these moments guide us to better designs. As a result, an iterative prototyping process is essential for communicating with the domain experts about all aspects of a design. As more and more capabilities are revealed to the domain experts, they are able to better understand the full potentials of different visualization techniques, thus leading to better design decisions. Our experience confirms the need of user-centered design. ``Too often, systems are designed with a focus on business goals, fancy features, and the technological capabilities of hardware or software tools. All of these approaches to system design omit the most important part of the process --- the end user.''~\cite{Labs_undated-dq}.


\section{Full Realization}
\label{sec:iterative_refinement_stage}

In this last stage, we pursued formal evaluations of our design. 
At this point a fully working prototype that meets all of the defined goals is complete. However, this stage focuses on identifying and correcting unforeseen problems or issues that only become apparent 
over actual usage of the data analysis support.   Others 
have argued that a valid evaluation of a design, say a software
tool, can only be achieved with unsupervised use by end users~\cite{Shneiderman2006-lq,Valiati2008-nf,Slingsby2012-gs}. 
Although this is true to some extent, we believe 
a carefully designed test 
can help us verify specific usability considerations, 
as demonstrated by others~\cite{Wu2009-js,Bateman2010-ai,Tory2014-km}. Many visualization researchers perform usability testings exclusively for the final evaluation of their system. However, we prefer 
to embed usability testing throughout the development process, 
as they allow us to identify usability problems early in the development process to improve our design.


Formal usability testing could demand a significant amount of 
the domain experts' time. To overcome this issue, Tory and Moller~\cite{Tory2005-ex} and Hearst et al.~\cite{Hearst2016-hb} show that using usability experts instead of domain experts can also discover major usability issues.
At the same time, we also believe that by involving more participants, more usability issues can be discovered, thus resulting in a better solution.
Consequently, we employ two types of usability testing, each with its own iterative loop. The first is \textit{User Experience Testing}, which focuses on discovering user experience issues by employing those who are not domain experts but are familiar with interactive visualization interface and scientific data analysis, 
Following the tests, necessary improvements and modifications are made to the software accordingly. 
Once this step is complete, we move onto \textit{Domain Expert Usability Testing}, which uses the actual users as test participants. 
This type of usability testing focuses on identifying domain-specific issues that could not be detected by a non-expert user, such as the ability to test hypotheses and gain new insights from the data. This testing should be repeated until the domain experts are satisfied with the product.



\subsection{User Experience Testing}
\label{sec:user_experience_testing}


For user experience testing, we mainly obtained qualitative results from observing user behaviors and asking open-ended questions.
The detailed design and result interpretation of the testing are presented as follows.

\subsubsection{Setup}

\noindent
\textbf{Participants.}
We recruited a total of 10 participants, who are all graduate
student researchers. 
Based on the self-reporting background questionnaire, 9 participants are familiar with data visualization, 7 participants are familiar with scientific visualization, 6 participants are familiar with 
scientific simulation data, and 9 participants are familiar with PDFs.

\noindent
\textbf{Apparatus.}
Our prototype is implemented in C++ and OpenGL using Qt as the GUI framework. 
The computer used for the evaluation is a 2013 MacBook Pro with a 13.3 inch retina display at 2560 $\times$ 1600 resolution. 
Users could choose to use the built-in trackpad or an external mouse.
We conducted the study in our lab, a quiet, office-like setting free from outside interruptions.
During the user study, screen recording with audio and interaction logging were used to capture the detailed behaviors of the participants. The test administrator sat alongside 
the participant to provide necessary guidance.

\begin{figure*}[t]
	\centering
	\begin{subfigure}[t]{0.357\textwidth}
		\includegraphics[width=\textwidth]{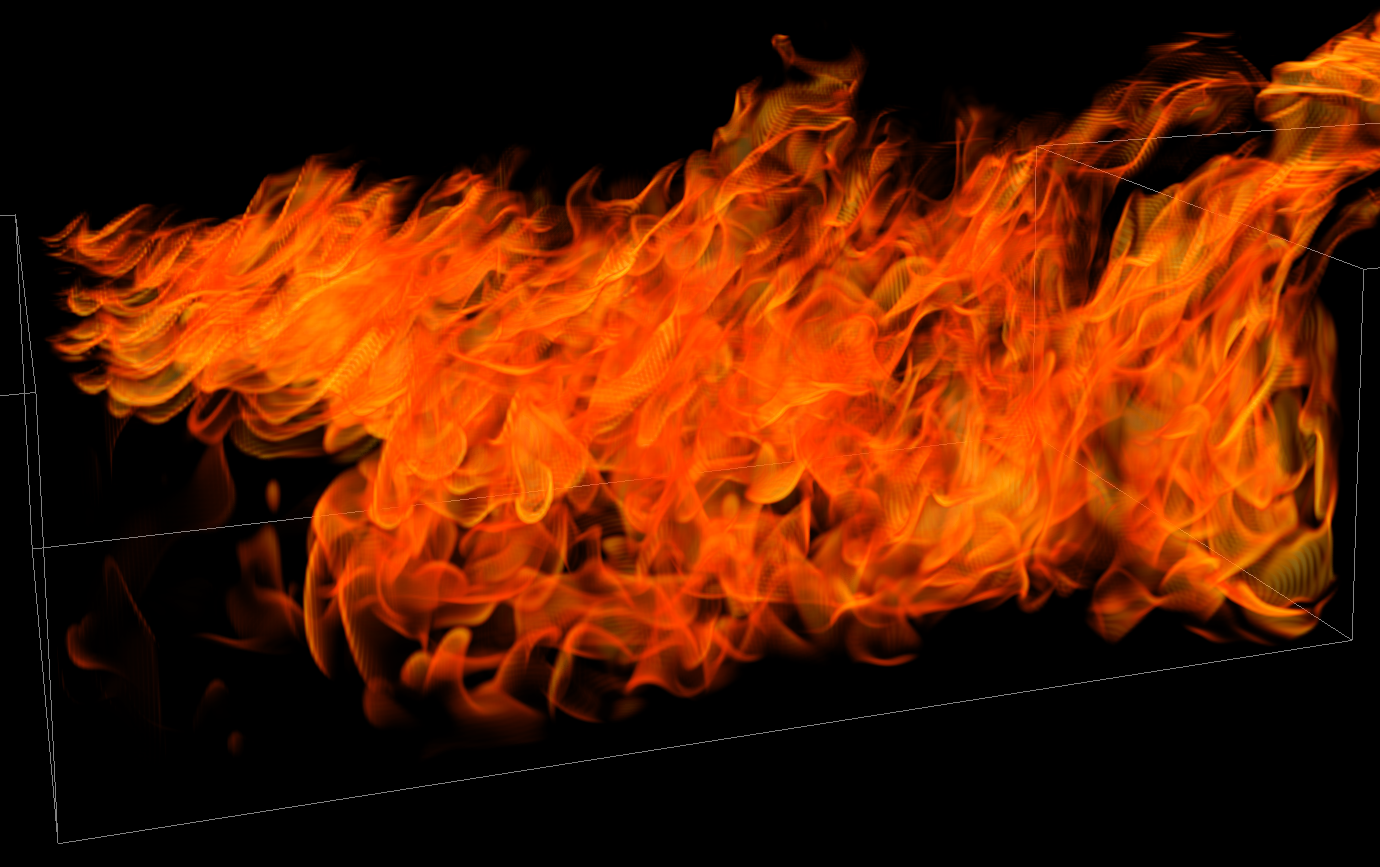}
		\caption{A Volume Rendering}
		\label{fig:cavity:volren}
	\end{subfigure}%
	\hspace{\fill}
	\begin{subfigure}[t]{0.402\textwidth}
		\includegraphics[width=\textwidth]{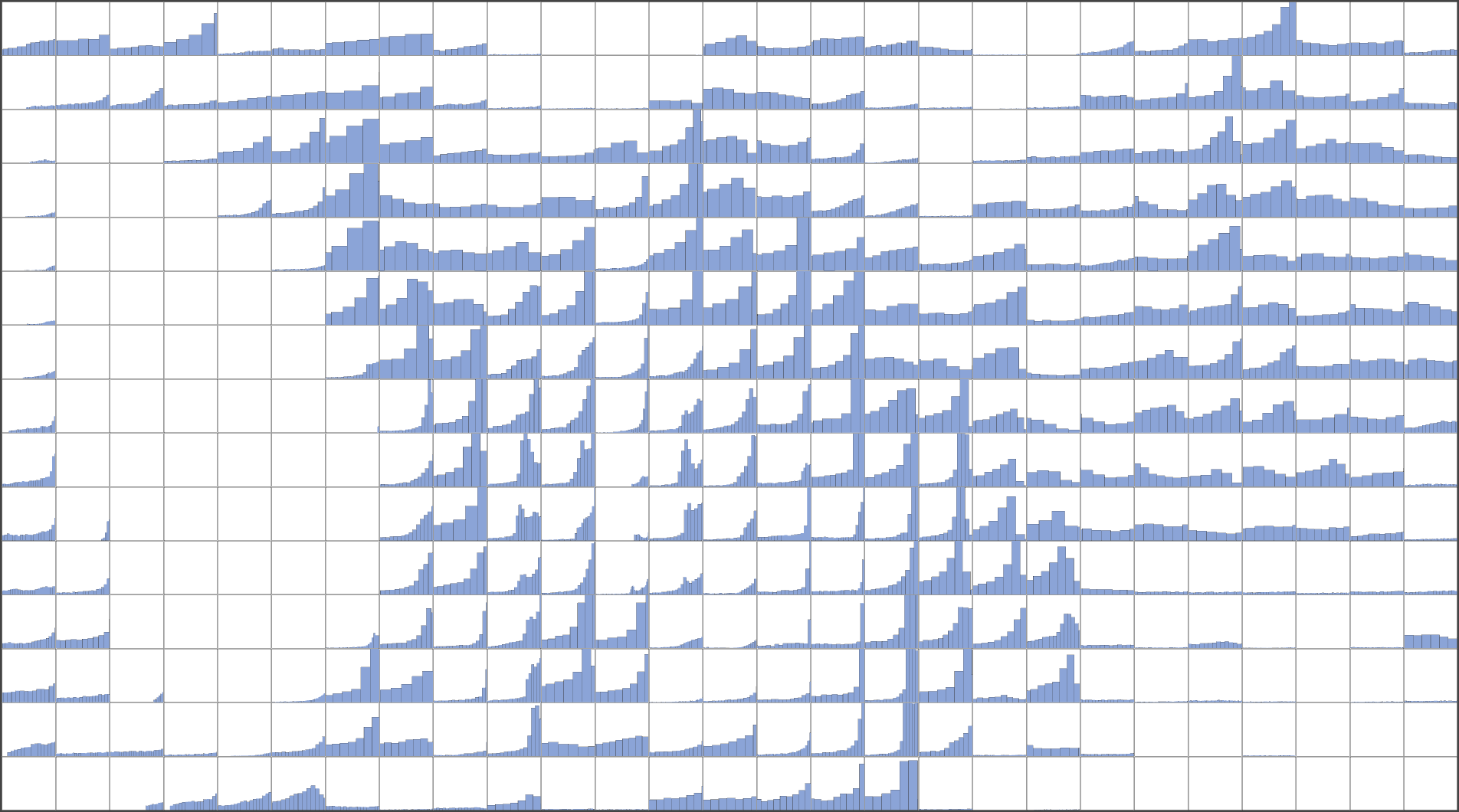}
		\caption{A Slice of 1D PDFs}
		\label{fig:cavity:pdf}
	\end{subfigure}%
	\hspace{\fill}
	\begin{subfigure}[t]{0.225\textwidth}
		\includegraphics[width=\textwidth]{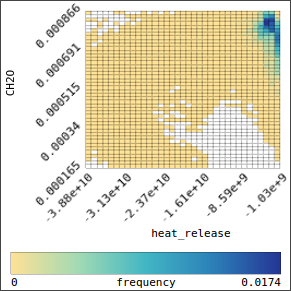}
		\caption{A 2D PDF}
		\label{fig:cavity:quenching}
	\end{subfigure}
	\caption{The \textit{cavity} dataset generated by a large-scale S3D simulation. (\protect\subref{fig:cavity:volren}) A direct volume rendering of heat release shows the cavity at the time step 5.2310E-04 of the simulation volume. (\protect\subref{fig:cavity:pdf}) A slice of the 1D PDFs in the cavity at the time step 5.2310E-04 depicting heat release in the range of {[}-2e+10, -1e+10{]}. It reveals the same spatial pattern as the volume rendering.
This is a possible strategy for the first task in Section~\ref{sec:ux:tasks}.
(\protect\subref{fig:cavity:quenching}) A 2D PDF of heat release and $CH_2O$ represents a region of quenching.
The blue colors at the top right corner represent a high concentration of $CH_2O$ and an absence of heat release. Note that the heat release values are negative thus confusing three of the domain experts initially.}
	\label{fig:cavity}
\end{figure*}

\noindent
\textbf{Dataset.}
We used the \textit{cavity} dataset~\cite{Rauch2018-mq} for the testing, which is a multivariate time-varying dataset generated by a large-scale combustion simulation. The simulation aims to explore the flame stabilization under the influence of a protruding cavity. A snapshot is shown in Figure~\ref{fig:cavity:volren}. For the
testing, both heat release and mass fraction of $CH_2O$, which are the key quantities in identifying a flame, were used to generate 
time-varying joint PDFs using the Titan supercomputer at Oak Ridge National Laboratory. The participants could freely explore both the joint (2D) PDFs (i.e., heatmaps of heat release and $CH_2O$) and the 1D PDFs (i.e., histograms of heat release or $CH_2O$).

\noindent
\textbf{Tasks.}
\label{sec:ux:tasks}
Since our system is designed for domain experts, the tasks must be carefully designed to balance completable actions while being comprehensible. With the help from the domain experts, we first identified the common tasks they want to perform with our system, such as ``classifying simulation subdomains based on the distribution of $CH_2O$.'' Then we stripped out the domain knowledge requirement from these tasks by replacing the domain specific terminologies with commonly understood language, such as ``selecting PDFs that represent high concentrations of fuel.'' Based on the identified tasks, we were able to classify the tasks into three categories. Then we chose five tasks for the user experience study while ensuring there is at least one task from each category.
The following are the tasks each participant went through in order and the corresponding categories:
\begin{enumerate}
\setlength\itemsep{0pt}
    \item {{Describe a trend:} The particular heat release range of interest is from -2e+10 to -1e+10. The samples in this range represent a burning region or a flame. Describe what is happening within this range spatially and temporally.}
    \item {{Classify regions:} At the first XY slice of time step 16, select all PDFs that have a similar pattern to the given PDF within the same variable ranges.}
    \item {{Interpret PDFs:} In the selected PDF, which single bin has the highest frequency? What is the bin frequency and the bin value range (for both heat release and $CH_2O$)?}
    \item {{Interpret PDFs:} Identify the PDF with the lowest heat release within the selected PDFs.}
    \item {{Interpret PDFs:} In the previously selected PDF, what is the frequency value within heat release range {[}-9e+9, -8e+9{]}?}
\end{enumerate}


\noindent
There are no strictly correct answers to the tasks. Instead, they were designed so that we can observe the thought process of the participants and learn about the strategies they used for each task. Such observations allowed us to identify hiccups in the user experience and evaluate the effectiveness of the system features.

\subsubsection{Procedure}

The following step by step procedure was performed in the user study:
\begin{enumerate}
\setlength\itemsep{0pt}
\item The participants are first introduced to the purpose of the study, which is gathering user experience feedback to improve the tool.
\item The participants then fill out a background questionnaire. The purpose is to understand the participants' level of expertise with data visualization and scientific data. More specifically, the knowledge of volumetric data and the experience to interpret PDFs.
\item A tutorial session is followed to teach the participants how to use the visualization tool. It explains the functionalities of all the views and controls in the user interface. The participants can try out the visualization tool as long as they want during the tutorial.
\item After the tutorial, three exercises are used to ensure the participants know how to use the visualization tool.
\item The participants are then asked to complete five tasks. These tasks are listed in Section~\ref{sec:ux:tasks}. 
\item Next, a post-survey interview is performed to gather usability information and general feedback.
\item Finally, the participants fill out a questionnaire. 
\end{enumerate}

\subsubsection{Measures}

As the purpose of this study is to gather user experience problems, we want to observe the thought processes of the participants and the strategies they use when they are working on the tasks. Thus, we record the screens and the voices of the participants. The qualitative results from the recordings not only allow us to identify user experience issues, but also allow us to discover common strategies, which can be used to further simplify the workflow.

On the other hand, we also want to gather quantitative results by logging user interactions, which include the usage frequency of each feature, the time it takes a participant to complete each task, etc. These logs give us a statistical view into how effective the features are.

In order to gather information about the overall user experience from the participants, a post-study interview was conducted with participant's voices recorded and a post-study questionnaire was filled out by the participants. The interview gather the participants' feedback in a conversational environment. The post-study questionnaire used Likert ratings to gauge the overall satisfaction level of the participants. Optional comments were also gathered using the questionnaire on individual features of our system.

\subsubsection{Results}

\begin{figure}[ht]
    \centering
    \includegraphics[width=\columnwidth]{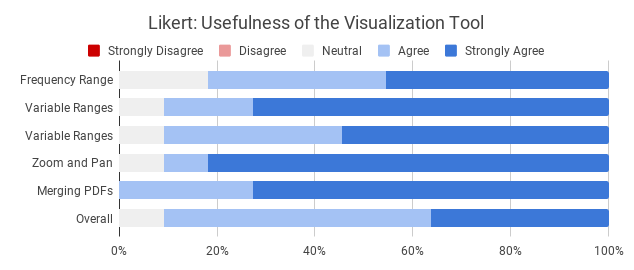}
    \caption{Results of the Likert scale ratings gathered in the post-study questionnaire. The ratings are overwhelmingly positive. Zooming and panning are among the highest rated features yet they are underutilized during the user experience testing.}
    \label{fig:ux-likert}
\end{figure}

As indicated by the Likert scale ratings (Figure~\ref{fig:ux-likert}), the participants believed the visualization tool would be useful for the domain experts. All tasks were finished with reasonable results by each participant. Most of the participants were able to immediately identify the required controls for the corresponding tasks. All user interface controls were utilized throughout the usability testing except the zoom and pan feature.
Some of the participants stated that the tool would feel more intuitive if they had more domain knowledge about the data. Many usability issues were also raised by the participants. The details of these issues, organized by user interface controls, are 
discussed as follows.


\noindent
\textbf{Variable Range Controls.}
\label{sec:ux-results:variable-range}
For the UI feature to modify variable ranges, the participants reported in the post-study questionnaire that the feature was useful for observing trends and patterns, as indicated at the second and third rows of Figure~\ref{fig:ux-likert}. Two participants commented that ``having text input boxes was useful to input accurate ranges'' as the numbers were usually too complicated to be input by sliders. Even though some participants verbally doubted whether the text boxes could accept scientific notation, all of them soon got to a positive conclusion. Three participants said allowing scientific notation would be helpful while one participant insisted that ``sliders would be much more intuitive for specifying ranges.''

In the PDF view, the users were able to brush to select the bins, and in turn, change the variable ranges. The selected variable ranges were shown in the variable range text boxes. However, the opposite did not hold true. When the users specified the variable ranges in the text boxes, the PDF view was not updated to indicate the new variable ranges. This behavior caused significant confusion to the participants, as eight participants reported that the PDF view should be synchronized to the variable range text boxes.
This observation suggests that when implementing the brushing and linking feature, the views should always be synchronized; 
otherwise, it would create tremendous confusion. This usability issue was not discovered by us during the development phase as we only expected the slice view to be the recipient of the change of variable ranges. It is easy to be tunnel visioned, and as a result usability studies are extremely important for user-driven development.\\

\noindent
\textbf{Spatial and Temporal Controls.}
\label{sec:ux-results:spatial-temporal}
One major usage of the visualization tool is to identify temporal and spatial trends and patterns in the data. All participants could easily observe and describe the patterns quickly and correctly using the slice view. All participants used the timeline view to explore the temporal dimension of the data.
Only one participant misinterpreted the horizontal axis of the slice view as the temporal axis while it represented one of the three-dimensional spatial axes, which was corrected by the administrator.
This participant self indicated as only familiar with information visualization. Five participants also utilized the slice index slider to compare the patterns in other slices of PDFs, while four of 
these five participants indicated that they were familiar with scientific visualization and volume data.


Another big surprise is that the zoom and pan feature were rarely used to navigate the slice view. The expectation was to allow participants to zoom in to analyze the PDFs in detail in the slice view. However, only four participants utilized the zoom and pan feature. One participant suggested that ``I generally didn't need to use the zooming and panning because the number of PDFs was small enough that I could see them all pretty well at once.'' Another participant suggested that the PDFs in the slice view did not show enough information (e.g axes and labels were not shown) to trigger users to zoom in. As a result, users tend to use the PDF view for detail analysis of individual PDFs. It is possible to also draw the axes and labels in the slice view when the PDFs are zoomed in.
A level-of-detail approach can be deployed.\\

\noindent
\textbf{Selection and Interpretation.}
\label{sec:ux-results:selection-interpretation}
The feature of switching between 1D and 2D PDFs was expected to be used extensively for reading the PDFs. 1D PDFs should be better at reading frequency values of a single variable while 2D PDFs could be used to look at the relationships between two variables. However, only three participants utilized this feature by themselves. The administrator taught the other participants when they failed to use this feature and the other participants stated that the feature was useful after they realized it. Therefore, one possible explanation was our system did not provide enough hints for the users to switch between the 1D and 2D PDFs. It might be due to the placement of the control element at the bottom of the control panel, which is sometimes hidden. Possible solutions include showing both the 1D and 2D PDFs in the same view or placing the control to switch between 1D and 2D PDFs to a more obvious place, i.e., inside the PDF view.


The PDF merging feature was used extensively in the study mainly because the participants were specifically asked to do so. All participants reported this feature to be useful in the post-study questionnaire, as shown in the forth row of Figure~\ref{fig:ux-likert}. However as indicated by one participant, ``It was useful for the task, but since I don't have domain knowledge, I do not know its utility in an actual analysis scenario.'' Another participant reported that ``I think this is most useful if you know how to find a feature of interest. For general exploration, though, I think it is less helpful.'' Another problem with the merging feature was the lack of uncertainty information arising from the merging algorithm. However, such uncertainty assessment is out of the scope of our study. 



\subsubsection{Changes and Improvements}

Before moving onto the domain expert usability study, we made the following improvements to the visualization tool based on the results from the user experience study:
\begin{itemize}
\setlength\itemsep{0pt}
\item Highlight the bins in the PDF view according to the variable range controls, as discussed in Section~\ref{sec:ux-results:variable-range}.
\item Utilize a level of detail technique to render the PDFs in the slice view, which progressively shows more information when zooming in, as discussed in Section~\ref{sec:ux-results:spatial-temporal}.
\item Add a small orientation view at the corner of the slice view to show where the current slice locates in the 3D volume, in the hope to clear the confusion discussed in Section~\ref{sec:ux-results:spatial-temporal}.
\item Clarify the frequency range controls and variable range controls by reorganizing the interface and replacing the labels, as discussed in Section~\ref{sec:ux-results:selection-interpretation}.
\item Show the aggregated statistics of the brushed bins in the PDF view, as requested by all participants.
\item Implement a lasso tool to select multiple PDFs in the slice view, as requested by five participants.
\item Automatically populate the frequency/variable range controls to eliminate the confusion arising from partially filled ranges, which was raised by three participants.
\item Provide a visualization of the overall statistics of each time step in the timeline view, as requested by one participant.
\end{itemize}

\subsection{Domain Expert Usability Testing}
\label{sec:domain_expert_usability_testing}

After enhancing the visualization tool based on the user experience testing, we evaluated how useful the entire solution is for the domain experts. Our solution includes an in situ library and a post hoc visualization tool. For the in situ library, we had conducted performance analyses to evaluate its impact to the S3D simulation~\cite{Ye:2016}. 
In this usability testing, our objective is to 
identify any difficulties over integrating the in situ 
library into their simulation with the provided documentation. 
For the post hoc visualization tool, we directly involved the domain experts as participants in another session of the usability testing. The testing for the post hoc visualization tool shared a great amount of similarity with the user experience testing. The apparatus used in both tests is the same. The measurements are also the same, which include screen capturing and interaction logging. 

\subsubsection{Participants and Tasks}

We were able to invite four domain experts to participate in the user study. One of the domain experts provided the data for the tasks, who is also a coauthor of this paper. All of them were familiar with the underlying scientific phenomenon of the data.

\begin{figure}[ht]
	\centering
	\includegraphics[width=\columnwidth]{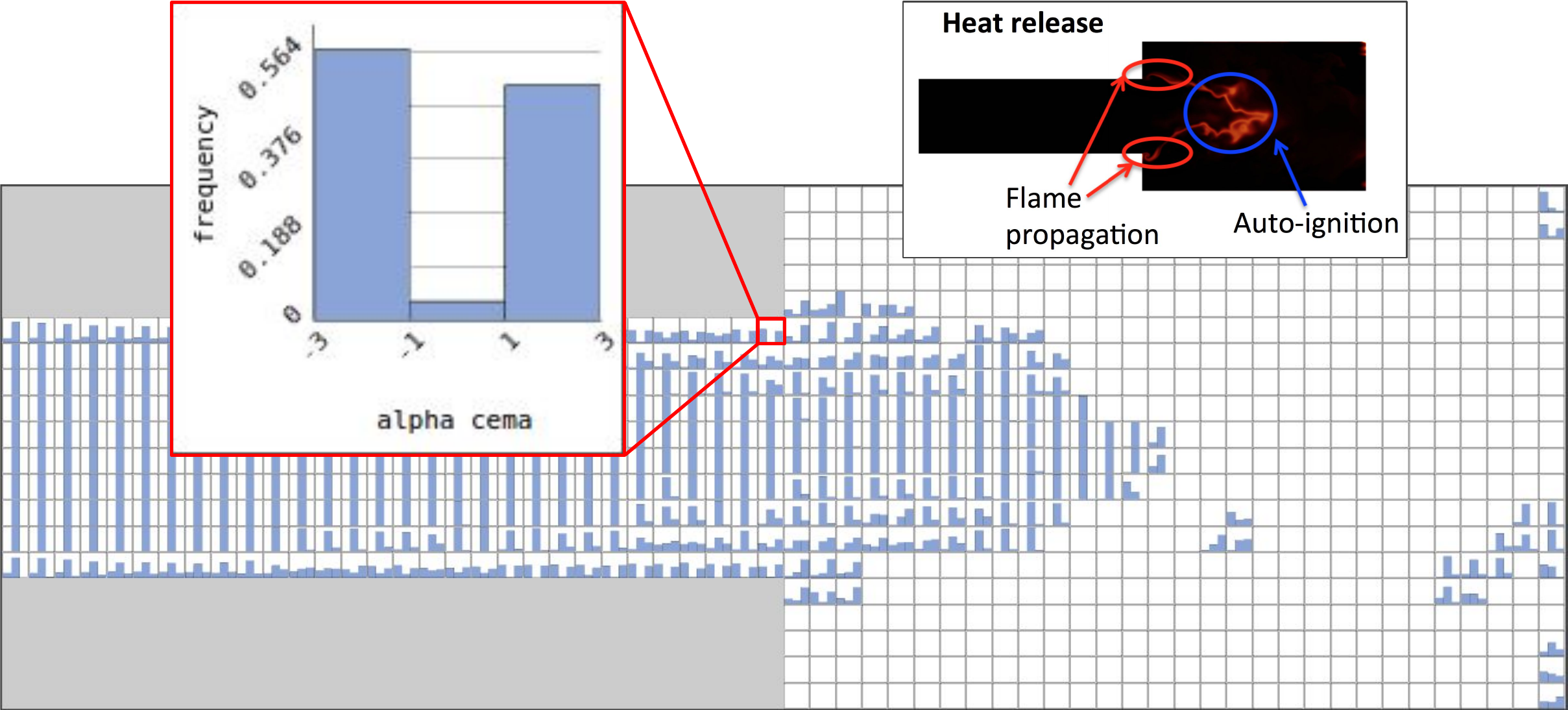}
	\caption{A slice of 1D PDFs is generated by conditionally binning the results of the CEMA analysis~\cite{Aditya2018-xd}. The shape of the slice matches the combustion chamber. The empty PDFs represent the regions with burnt products of combustion, where there no significant reactions. Zoomed PDF: the high frequencies in both the left bin and the right bin indicate that this region represents the propensity of the fuel-air mixture to burn under flame propagation. The left bin represents samples in extinction zones and the right bin represents samples in assisted ignition zones.}
	\label{fig:cema}
\end{figure}

The tasks were designed to require domain knowledge to complete.
Also, the tasks should demonstrate how the visualization tool can help the domain scientists to efficiently explore the scientific phenomenon. Finally, we decided to duplicate one task from the user experience testing, with the hope to observe the different user behaviors between visualization experts and domain experts. The three tasks are:
\begin{enumerate}
\setlength\itemsep{0pt}
\item{{\textbf{Duplicated Task.}} \textit{The particular heat release range of interest is from -2e+10 to -1e+10. The samples in this range represents a flame is burning in the particular region. Describe what is happening within this range spatially and temporally.} It is expected that the domain experts can rely on their domain knowledge to dig out deeper information.}

\item{{\textbf{Cavity Task.}} \textit{Select PDFs that represent regions of quenching (where the flame gets extinguished due to the heat losses) at the time step 6.3110E-04. Quenching typically occurs in the near-wall regions. In this case, the quenching events can be identified with significant presence of $CH_2O$ and absence of heat release.} A sample PDF representing a region of quenching is shown in Figure~\ref{fig:cavity:quenching}.}

\item{{\textbf{CEMA Task.}}  \textit{Select subdomains where flame propagation mode is likely to be present in a reheat burner dataset. This task uses the data quantified by the chemical explosive mode analysis (CEMA)~\cite{Aditya2018-xd}. The data can be understood as binning ``alpha'' values to classify the combustion modes (auto ignition and flame propagation).} The flame propagation mode can occurs in the regions of assisted ignition or extinction zone, which are based on the range of $alpha$. This task utilizes the data from a reheat burner simulation~\cite{Aditya2018-xd}, which is processed into conditional PDFs in situ using the Titan supercomputer. A slice of the PDFs and a sample PDF representing flame propagation is shown in Figure~\ref{fig:cema}.}
\end{enumerate}

\subsubsection{Results}

For the session to integrate the in situ library, the domain experts spent approximately one hour, while less than ten minutes were spent integrating and debugging the in situ routines. Much of the remaining time was spent discussing the details of the in situ library, including the underlying algorithms, performance impacts to the S3D simulation, etc. As we explained the algorithms, showed the results from the software test suite, and presented the performance comparison results, they were convinced the in situ library was ready to be integrated into a production simulation code.


As to the testing for evaluating the post hoc visualization tool, the domain experts gave exceptionally positive responses. Due to the small number of participants, most results were qualitative. They liked that the visualization tool was able to give an overview of the data while also providing 
certain statistical analysis results.\\ 

\noindent
{\bf Applying Domain Knowledge.} 
Less time was spent on teaching the domain experts how to use the visualization tool, and the domain experts used less time to complete the exercises. Also, while the visualization experts doubted whether scientific notation was accepted by the text boxes, the domain experts did not question it at all. We believe the reason is twofold: many usability issues were fixed after the user experience testing and the domain experts were more familiar with statistical analysis. 

One participant ventured into his/her own journey of analyzing the data after the tutorial session, which resulted in he/she accidentally completing one of the tasks before the task session. The administrator did not stop this ``adventure'' as we believed it was a good sign that the visualization tool inspired the participant to explore the data.

For the duplicated task, which was the task to describe the trend spatially and temporally, the domain experts performed as well as the usability experts. The difference was that the domain experts applied domain knowledge to explain the observed patterns, as one participant tried to explain to the administrator: ``Eventually, everything burns up here, so now it's generating a flame that's going out of the cavity.'' One participant also doubted that the heat release range we provided in the task was precise enough, but he/she agreed that it was a reasonable range for the usability testing. We were happy to observe such phenomenon as it indicated that the domain experts were able to verify their domain knowledge with the visualization tool.

For the cavity task, the domain experts were able to identify the quenching regions after a small confusion was cleared up. They assumed the right side of a PDF corresponds to higher values of heat release. However, the heat release data were provided as negative values, thus the left side of a PDF means higher values of heat release. Therefore, the flipped PDFs confused three of the domain experts. After explaining the negativity of the heat release variable, they were able to realize the horizontally flipped patterns and correctly classify the quenching regions. The only domain expert who was not confused was the author of the data.


For the CEMA task, one participant completed the task using the temperature PDFs while the expected strategy was to use the PDFs from the CEMA results. Despite the different strategies, the chosen PDFs were mostly correct. After the administrator questioned about the participant's strategy and explained the intended usage of the CEMA PDFs, the participant admitted that he/she was not familiar with the CEMA technique and explained that ``based on the PDFs, portions where these peaky temperature are toward the higher end is where flame propagation usually is.'' We were happy to observe that our visualization tool was flexible enough for the domain experts to perform analysis with different strategies.\\

\noindent
{\bf Usability Issues and Feature Requests.}  
The feature the domain experts liked the most was the capability to merge PDFs, as indicated in the post questionnaire and verbally, ``It's nice we can aggregate statistics.'' One participant was excited about the feature and tried to merge all PDFs in the current slice to obtain the aggregated statistics. The operation was expensive thus rendering the user interface irresponsive for a couple of minutes. This could be overcome by optimizing the merging algorithm and precalculating the aggregated statistics. Another issue raised by the domain experts was that the feature could only provide an estimation because the bin ranges of the PDFs can be different. They also suggested that quantifying the uncertainty of the algorithm could make the feature more trustworthy.

Another user experience issue was raised by one participant as he/she was confused by the horizontal and vertical axes of the slice view. The 3D orientation widget provided the location of the current slice in the simulation volume, but failed to indicate which axis of the current slice corresponds to which axis in the volume. An extra 2D orientation widget could be used to solve this issue.

Multiple feature requests were also placed by the domain experts:
\begin{itemize}
\setlength\itemsep{0pt}
    \item Three domain experts suggested the visualization tool could provide calculations of high level statistics, such as means and variances. One participant suggested that the logical questions to ask after identifying a region of interest was ``what is the probability of my variable being within a range or what is the joint probability of variable one and variable two?'' The participant also reported that ``we start with visually interesting portions of the domain, then we actually report hard numbers.''
    \item Two participants wanted to overlay the PDF slice with an image of the slice as the background. They believed such setup could help them locate the PDFs in the simulation volume.
    \item One domain expert requested to export the merged PDF so that they could perform further analysis with other tools.
    \item All participants liked the statistics shown on the time step slider. One participant indicated interest in having greater control of the statistics shown.
\end{itemize}

\section{Conclusion}


We successfully applied user-centered design principles 
to the development of scientific visualization solutions 
for domain experts. We took lessons from 
 information visualization work as well as 
our own experience through a long term collaboration 
 with scientists in realizing in situ visualization and  distribution-based data analysis. Our design process focuses on communicating with the domain experts and closing the domain knowledge gap, over three stages: domain analysis, iterative development, and iterative refinement. The experiences gained and lessons learned from each stage of the collaboration 
 are valueable. 
We hope that this work inspires the scientific visualization community to more commonly adopt user-centered design practices as it is crucial to developing truly usable solutions.

For future work, we expect that many additional in situ data visualization and analysis methods will be added. 
Therefore, our iterative design process will continue. 
One feature of high interest is to provide the uncertainty information of merged PDFs. The domain experts expressed interest in the interactive PDF merging feature and we see great potential of this functionality. However, since the bin ranges of the PDFs can be different and the underlying samples are not present in post processing, the merging is a lossy operation. To better inform the domain experts about the statistics in the selected regions, we 
need to quantify and visualize the uncertainty information of the merged PDF.

Another desire capability is real-time monitoring of 
the simulation using PDFs. Such functionality allows domain 
experts to find abnormal activities of a running simulation by using statistical analysis. The PDFs are suitable for this task due to the small data size and the statistical analysis they enable. To implement such capability, the in situ generated PDFs needs to be streamed from a supercomputer to a local workstation, which is then rendered by the visualization tool.

Our experience indicates that utilizing visualization or 
HCI experts to identify user experience issues is an effective method. We were able to observe similar patterns as indicated in previous works~\cite{Tory2005-ex,Hearst2016-hb}. In particular, we agree with Tory and Moller~\cite{Tory2005-ex} that ``while expert reviews can provide quick and valuable insight into usability problems, they should not be used exclusively and should not replace user studies.''





\acknowledgments{
This research is sponsored in part by the U.S. Department of Energy through grant DE-SC0012610 and DE-SC0019486.
In situ processing work used the Oak Ridge Leadership Computing Facility. 
}

\bibliographystyle{abbrv-doi}

\bibliography{template}

\end{document}